\documentclass[12pt]{article}
\usepackage{amsfonts,latexsym}
\begin{document}
\newcommand{\e}{\epsilon} \newcommand{\ot}{\otimes}
\newcommand{\be}{\begin{equation}} \newcommand{\ee}{\end{equation}}
\newcommand{\ba}{\begin{eqnarray}} \newcommand{\ea}{\end{eqnarray}}
\newcommand{\tmod}{{\cal T}}\newcommand{\amod}{{\cal A}}
\newcommand{\bemod}{{\cal B}}\newcommand{\cmod}{{\cal C}}
\newcommand{\dmod}{{\cal D}}\newcommand{\hmod}{{\cal H}}
\newcommand{\s}{\scriptstyle}\newcommand{\tr}{{\rm tr}}
\newcommand{\einsop}{{\bf 1}}
\def\oR{R^*} \def\upa{\uparrow}
\def\R{\overline{R}} \def\doa{\downarrow}
\def\nn{\nonumber} \def\dag{\dagger}
\def\be{\begin{equation}}
\def\ee{\end{equation}} 
\def\bea{\begin{eqnarray}} 
\def\eea{\end{eqnarray}} 
\def\ve{\epsilon}
\def\si{\sigma}
\def\th{\theta} 
\def\a{\alpha} 
\def\b{\beta}
\def\g{\gamma}
\def\h{\overline{h}}
\def\d{\delta}
\def\m{\eta} 
\def\n{\tau}
\newcommand{\reff}[1]{eq.~(\ref{#1})}
 \centerline{\bf{A CONSTRUCTION FOR R-MATRICES WITHOUT}} 
\centerline{\bf{ DIFFERENCE PROPERTY IN THE SPECTRAL PARAMETER }} 
~~~\\
\begin{center}
{\large Jon Links}
\vspace{1cm}~~\\
{\em Centre for Mathematical Physics, \\ Department of Mathematics, \\
The University of Queensland,  4072 \\  Australia \\
e-mail jrl@maths.uq.edu.au}
~~\\
~~\\
\end{center} 
~~\\
\begin{center}{Report No. UQCMP-99-2}
\end{center}
\vspace{1cm}~~\\
\begin{abstract}
A new construction is given for obtaining $R$-matrices which solve the 
McGuire-Yang-Baxter equation in such a way that the spectral parameters
do not possess the difference property. A discussion of the derivation of
the supersymmetric $U$ model for correlated electrons is given in this context,
such that applied chemical potential and magentic field terms can be 
coupled arbitrarily. As a limiting case the Bariev model is obtained.
\end{abstract}
\vspace{1cm}
\begin{flushleft}  
\end{flushleft}

\vfil\eject
\centerline{{\bf 1. Introduction}}
~~\\

The study of integrable models which are constructed through the Quantum 
Inverse Scattering Method (QISM) makes available many significant 
and non-perturbative results as 
these models are generally solvable through Bethe  ansatz methods.   
The quantum algebras \cite{j1,d1} and their natural $\mathbb 
Z_2$-graded analogues
 quantum superalgebras \cite{y,man} provide us with the principle examples of
quasi-triangular Hopf (super)algebras which allow us to systematically construct
solutions of the McGuire-Yang-Baxter (MYB) equation which are
 central to the QISM approach. 
The solutions of the MYB equation are referred to 
as $R$-matrices. The approach of the QISM is to first build a family of 
commuting transfer matrices from the $R$-matrices 
and then define the Hamiltonian operator as the
logarithmic derivative of the transfer matrix 
evaluated at the shift point. The shift point is 
that value of the spectral parameter such that the $R$-matrix behaves as the 
permutation operator. Those $R$-matrices that do have a shift point are 
called {\it regular} and it is this property which ensures that this
 definition gives a global Hamiltonian which is a sum of local 
two-site Hamiltonians.

The structure of the affine quantum (super)algebras is such that
solutions of the MYB equation
obtained automatically possess the difference property in
the spectral parameter as a result of solving Jimbo's equations
\cite{j2,bgz}.
On the otherhand, there exist other solutions of the MYB equation which do 
not have the difference property such as Shastry's solution for the Hubbard 
model \cite{s}. Another model which has been derived through the QISM
is the Bariev model \cite{b} which was introduced as a solvable generalization
of Hirsch's hole superconductivity model \cite{h}. In independent 
works by Zhou \cite{z}
and Shiroishi and Wadati \cite{sw} integrability of the Bariev model was 
established by obtaining an appropriate solution of the MYB equation. Though
the solutions of \cite{z} and \cite{sw} are different, they 
both share the feature
that they do not have the difference property.  Energy spectra 
obtained through the 
algebraic Bethe ansatz have been studied in \cite{z1,mr}.

Yet another model which is of interest in the study of correlated electron 
systems is the supersymmetric $U$ model \cite{bglz}
and importantly in the context of the 
work discussed here its anisotropic generalization \cite{bkz,ghlz} 
which can be derived
in the framework of the QISM using an $R$-matrix associated with one-parameter
family of minimal typical representations of the quantum superalgebra
$U_q(gl(2|1)$. As such, the $R$-matrix used to derive this model 
does have the difference property. In a particular limit, the anisotropic 
supersymmetric $U$ model reduces to the Bariev model with the addition of a 
divergent 
chemical potential term. The fact that these two models are related yet 
have been individually derived from $R$-matrices of differing character 
  with respect to their spectral parameter dependence 
has been  somewhat mysterious.

In this paper a construction which relates models which differ only by 
field 
terms such as the chemical potential will be given and a connection between
$R$-matrices of the difference property type and those without the difference
property will be established. The connection lies in the use of twisting of 
the algebraic $R$-matrices by a suitable twistor which retains
the quasi-triangular Hopf-algebra. The twisting construction was developed in 
\cite{d2} and gave rise to the more 
general notion of quasi-triangular quasi-Hopf
algebras. However, provided the twistor satisfies appropriate properties 
known as the twisted 2-cocycle condition,
 the twisted structure can also be of the  Hopf algebra type. Such twisting
 operations have lead to some significant developments such as the construction 
of elliptic solutions of the MYB equation \cite{jkos,zg} 
and the construction of 
multiparameteric integrable systems \cite{flr}. The latter were based on 
a particular type of twistor due to Reshetikhin \cite{r}. Here, a generalized
form will be used which was gven by Engeldinger and Kempf \cite{ek}. By a 
certain parametrization, the regularity property of the $R$-matrix will be 
shown to still hold which  allows the construction of an integrable 
one-dimensional model with two-site interactions. This type of 
parametrization is adopted from the method used in \cite{l} to obtain
 an extended region of integrability for the supersymmetric $U$ model.

As an example, the aniostropic supersymmetric $U$ model will be constructed 
with arbitrary chemical potential and magentic field term. In an appropriate
limit the Bariev model is recovered. However, an attempt to obtain an $R$-matrix
solution for the Bariev model proves fruitless and still leaves open the 
question of the origins of the solutions obtained in \cite{z,sw} in terms of 
an underlying algebraic structure.
  
Finally, it worthwhile to observe that the construction employed here may 
be understood in terms of representations of {\it coloured} quantum algebras
\cite{q}. (These are not to be confused with {\it colour} algebras in the sense 
of \cite{mac}.) In fact, the relation between coloured quantum algebras and 
twisting procedures has been discussed by Chakrabarti and Jagannathan
\cite{cj}, however it appears that their role in relation to integrable 
systems has not yet been addressed in the literature.     
\vspace{1cm}
~~\\  
\centerline{{\bf 2. Quantum Inverse Scattering Method }}
~\\
Let $R(u),\,\R(u,v)\in \mathrm{End}\,V\ot V$ give a solution of the
 MYB equation
\be 
\R_{12}(u,v)R_{13}(u)R_{23}(v)=R_{23}(v)R_{13}(u)\R_{12}(u,v). \label{yb}
\ee 
For full generality, we consider the cases when 
 $V$ denotes  a ${\mathbb Z}_2$-graded vector space. In such instances
it is necessary to impose the following rule for the tensor product
multiplication of matrices:
\be (a\ot b)(c\ot d)=(-1)^{(b)(c)}ac\ot bd \label{rule}\ee 
for matrices $a,\,b,\,c,\,d$ of homogeneous degree. The symbol $(a)\in
{\mathbb Z}_2$ denotes the degree of the matrix $a$. 
The transfer matrix is defined
$$t(u)=\mathrm{str}_0\left(R_{0N}(u)R_{0(N-1)}(u)....R_{01}(u)\right) $$
which from (\ref{yb}) can be shown to satisfy
$$[t(u),\,t(v)]=0,~~~\forall\,u,v\in {\mathbb C}.$$
Above $\mathrm{str_0}$ denotes the supertrace taken over the auxiliary
space which is labelled by 0.

In the usual manner the Hamiltonian associated with the transfer matrix
is defined by the relation
$$H=\left.t^{-1}(u)\frac{d}{du}t(u)\right|_{u=0}.$$
Assuming regularity of the $R$-matrix; i.e.
$$R(0)=P$$
where $P$ is the (${\mathbb Z}_2$-graded) permutation operator,  yields
\be H_{global}=\sum_{i=1}^{N-1} H_{i(i+1)} + H_{N1}
\label{global}\ee 
where the local two site Hamiltonians are given by
\be H=\left.\frac{d}{du}PR(u)\right|_{u=0}.\label{local} \ee

\vspace{1cm}

\centerline{{\bf 2. Solutions arising from quantum superalgebras}}
~~\\
Let $G$ denote a simple Lie (super)algebra of rank $r$ with generators
$\{e_l, f_l, h_l\}_{l=0}^r$ and let $\alpha_l$ be its simple roots. Here we 
adopt the convention that in the distinguished root basis $\a_0$ 
labels the unique odd simple root. The quantum
(super)algebra $U_q(G)$ can be defined with the structure of a
(${\bf Z}_2$-graded) quasi-triangular Hopf algebra \cite{gzb}.
We will not give the full
defining relations of $U_q(G)$ here (see e.g \cite{y,man}) 
but mention that $U_q(G)$ has a
coproduct structure given by
\begin{equation}
\Delta(h_l)=I\otimes h_l+h_l\otimes I\,,~~~\Delta(a)=a\otimes
  q^{-h_l/2}+q^{h_l/2}\otimes a\,,~~~a=e_l, f_l.
\end{equation}

Let $\pi$ be an irreducible representation 
(irrep) of $U_q(G)$ afforded by
the irreducible module $V(\Lambda)$ where $\Lambda$ denotes the highest
weight.   
Assume  that the irrep $\pi$
is affinizable, i.e. it can be extended to an irrep of the corresponding
quantum affine (super)algebra $U_q(\hat{G})$. Consider an operator
$R(x)\in {\rm End}(V\otimes V)$,
where $x\in {\bf C}$ is the multiplicative spectral parameter. 
It has been
shown by Jimbo \cite{j2} and also Zhang et. al. \cite{bgz} 
for the $\mathbb Z_2$-graded
case that a solution to the linear equations
\begin{eqnarray}
&&R(x)\Delta(a)=\bar{\Delta}
  (a)R(x)\,,~~~\forall a\in U_q(G),\nonumber\\
&&R(x)\left (x\pi(e_{\Psi})\otimes \pi(q^{-h_{{\Psi}/2}})+
  \pi(q^{h_{{\Psi}/2}})\otimes \pi(e_{\Psi})\right )\nonumber\\
&&~~~~~~  =\left (x\pi(e_{\Psi})\otimes \pi(q^{h_{\Psi}/2})
  +\pi(q^{-h_{\Psi}/2})\otimes \pi(e_{\Psi})\right )R
  (x)\label{r(x)1}
\end{eqnarray}
satisfies the Yang-Baxter equation
in the tensor product module $V\otimes V\otimes V$:
\begin{equation}
R_{12}(x)R_{13}(xy)R_{23}(y)
  =R_{23}(y)R_{13}(xy)R_{12}(x).
\end{equation}
In the above,
$\bar{\Delta}=T\cdot \Delta,$ 
with $T$ the twist map defined by
$$T(a\otimes b)=(-1)^{[a][b]}b\otimes a\,,~\forall a,b\in U_q(G)$$ 
and $\pi(e_{\Psi}),\,\pi(h_{\Psi})$ are matrices for operators associated 
with the highest root $\Psi$ such that the representation extends to a loop 
representation of the untwisted affine extension of $U_q(G)$.
The multiplicative
spectral parameter $x$ can be transformed into an additive spectral
parameter $u$ by $x=\mbox{exp}(u)$.
The structure of the Jimbo equations is such that the solutions obtained 
must necessarily have the difference property; i.e.
$$\R(u,v)=R(u-v).$$ 
Explicit solutions of the Jimbo equations can be computed by tensor 
product graph methods as discussed in \cite{zgb,dgz,dglz}.

As consequence of the Jimbo equations, for any Cartan element $h$ the operator 
$\sum_i h_i$ on the periodic lattice 
commutes with the global Hamiltonian (\ref{global}) and so we may 
arbitrarily add such terms to the Hamiltonian and still yield a solvable model.
In the next section, it will be described how the resultant model may 
be derived directly from an $R$-matrix without the difference property.

\vspace{1cm}

\centerline{{\bf 3. Twisting construction}} 
~~\\
Let $(A,~\Delta,~R)$ denote a quasitriangular Hopf (super)algebra
where
$\Delta$ and $R$ denote the co-product and $R$-matrix respectively.
Suppose that there exists an element $F\in A\otimes A$ such that
\begin{eqnarray}
&&(\Delta \otimes I)(F)=F_{13}F_{23},~~~~~~\nn  \\
&&(I\otimes \Delta)(F)=F_{13}F_{12},   \nonumber    \\
&&F_{12}F_{13}F_{23}=F_{23}F_{13}F_{12}. 
\label{fyb}
\end{eqnarray}
Then $(A,~\Delta^{F},~R^F)$ is also a
quasitriangular Hopf (super)algebra with co-product 
and $R$-matrix respectively
given by
\begin{equation}
\Delta^F=F_{12}\Delta F_{12}^{-1},~~~~~~~~R^F=F_{21}RF_{12}^{-1}.
\label{df}
\end{equation}    
Throughout we refer to $F$ as a {\it twistor}.

The result stated above is a little more general than that 
originally proposed by Reshetikhin and is due to Engeldinger and Kempf 
\cite{ek}. In the original work \cite{r} Reshetikhin 
imposed the additional constraint 
$$F_{12}F_{21}=I\otimes I $$ 
and in the case that $(A,~\Delta,~R)$ is an affine quantum (super)algebra
Reshetikhin gave the example  that $F$ can be chosen to be
\begin{equation}
F={\rm exp} \sum_{i<j}\left(h_i\otimes h_j-h_j\otimes
h_i\right)\phi_{ij}
\label{F}  \end{equation}
where $\{h_i\}$ is a basis for the Cartan subalgebra of the affine
quantum (super)algebra and the $\phi_{ij},~i<j$ are arbitrary complex
parameters. However following the construction of Engeldinger and Kempf 
it is possible to choose 
\begin{equation}
F={\rm exp} \sum_{i,j}\left(H_i\otimes H_j\right)\phi_{ij} 
\label{tw} \end{equation} 
which obviously gives a twistor dependent on more free parameters. 
For our purposes either approach may be adopted but we will choose the 
latter for convenience. 
Note also that it is also possible (and for the construction below 
essential) to extend the Cartan subalgebra by
an
additional central extension (not the usual central charge) $\h$
which
will act as a scalar multiple of the identity operator in any
representation.

It is worth observing here that the class of twistors 
(\ref{tw}) above also qualify as Hopf algebra preserving twistors of 
the type described by Andrews and Cornwell \cite{jc}
in their work on relating non-standard
quantum algebras to standard ones.
In notation as above, suppose there exists $F\in A \otimes A$
which satisfies the following relations
\bea F_{12}F_{23}&=&F_{23}F_{12} \nn \\
\left(\Delta \otimes I\right)F&=&F_{23}F_{13} \nn \\
\left(I\otimes \Delta\right)F&=&F_{12}F_{13}. \nn
\eea
Then  $(A,~\Delta^{F},~R^F)$ is also a quasi-triangular Hopf
algebra with $\Delta^{F},~R^F$ given by \reff{df} above.

For a given element $h$ of the Cartan subalgebra we now choose a twistor in
the particular form
$$F=\exp(\m h\ot \h+\n \h\ot h)$$
where $\m,\,\n $ are arbitrary scalars. This gives the twisted algebraic 
$R$-matrix the form
$$R^F=\exp(\m \h\ot h+\n h\ot \h)R\exp(-\m h\ot \h-\n \h\ot h)$$
which still satisfies the MYB equation
$$R_{12}R_{13}R_{23}=R_{23}R_{13}R_{12}.$$

Suppose again that $\pi$ extends to a loop representation of the affine quantum
superalgebra. We let $\pi\ot\pi\ot\pi$ act on the above equation but let the 
central element $\h$ take different values $\b,\,\g,\,\d$ in each of 
the tensor product spaces. Letting $R^F(u,\b,\g)$ denote the image of $R^F$
under $\pi \ot \pi$ where $h_0$ acts as $\b$ in the first space and $\g$
in the second, this then yields the matrix solution 
\be R_{12}(u-v,\b,\g)R_{13}(u,\b,\d)R_{23}(v,\g,\d)
=R_{23}(v,\g,\d)R_{13}(u,\b,\d)R_{12}(u-v,\b,\g) \label{nodiffmyb}\ee
>From this solution we may construct the transfer matrix
\be
t(u,\b,\d)={\rm str }_0\left( R_{0L}(u,\b,\d)....R_{02}(u,\b,\d)
R_{01}(u,\b,\d)\right)
\ee
which as a result of (\ref{nodiffmyb}) forms a commuting family in two variables
\be
[t(u,\b,\d), t(v,\g,\d)]=0.  \ee
It then follows that one may diagonalize $t(u,\b,\d)$ independently of 
both $u$ and $\b$.
Setting $\exp h=M$ and $R(u)$ as the image of $R$ under $\pi\ot \pi$ we get
$$R^F(u,\b,\d)=M_1^{\n\d}M_2^{\m\b}R(u)M_1^{-\m\d}M_2^{-\n\b}.$$

The above solution $ R^F(u,\b,\d)$ can be made regular by setting $\d=0,\, 
\b=u.$ The local two-site Hamiltonian operator associated with this solution
is given by
\bea H^F&=&\left.\frac{d}{du}PR^F(u,\b=u,\d=0)\right|_{u=0}\nn \\
&=&H+\m h_1-\n h_2  
\eea
with $H$ given by (\ref{local}) above. It immediately follows that the 
global Hamiltonians are related by 
$$H_{global}^F=H_{global}+(\m-\n)\sum_{i=1}^Lh_i.$$

\vspace{1cm} 
\centerline{\bf 5. The supersymmetric $U$ model}
~~\\
The quantum superalgebra $U_q(sl(2|1))$ has simple generators $\{e_0,f_0,h_0
,e_1,f_1,h_1\}$ associated with the Cartan matrix
$$A=\pmatrix{0&1 \cr -1&2}.$$
This algebra admits a non-trivial one parameter family of four-dimensional
representations which we label $\pi$ given by 
\bea \pi(e_0)&=&\sqrt{[\a]}e^1_2+\sqrt{[\a+1]}e^3_4 \nn \\
     \pi(f_0)&=&\sqrt{[\a]}e^2_1+\sqrt{[\a+1]}e^4_3 \nn \\
     \pi(h_0)&=&\a (e^1_1+e^2_2)+(\a+1)(e^3_3+e_4^4) \nn \\
     \pi(e_1)&=&-e^2_3 \nn \\
     \pi(f_1)&=&-e^3_2 \nn  \\
     \pi(h_1)&=&e^2_2-e^3_3. \label{rep} \eea
Above the indices of the elementary matrices $e^i_j$ carry the 
${\mathbb Z}_2$-grading $(1)=(4)=0,\,(2)=(3)=1$ and we employ the notation
$$[x]=\frac{q^x-q^{-x}}{q-q^{-1}}.$$ 
Associated with this representation  there is  a solution of the 
MYB equation with the difference property which is obtained 
by solving Jimbo's equations. 
The problem of obtaining this solution was considered in \cite{ghlz,bdgz} 
(see also
\cite{a}). Adopting the prescription detailed in the previous section, the 
following solution of (\ref{nodiffmyb}) is obtained with the choice
$$h=(\a+1)I-h_0-kh_1$$
where $k$ is a free variable.

\begin{footnotesize}
\bea &&R(u,\b,\d)= \nn \\
&&\pmatrix{ 
R^{11}_{11}&0&0&0&|&0&0&0&0&|&0&0&0&0&|&0&0&0&0 \cr
0&R^{12}_{12}&0&0&|&R^{12}_{21}&0&0&0&|&0&0&0&0&|&0&0&0&0 \cr 
0&0&R^{13}_{13}&0&|&0&0&0&0&|&R^{13}_{31}&0&0&0&|&0&0&0&0 \cr
0&0&0&R^{14}_{14}&|&0&0&R^{14}_{23}&0&|&0&R^{14}_{32}&0&0&|&R^{14}_{41}&0&0&0 
\cr
-&-&-&-&|&-&-&-&-&|&-&-&-&-&|&-&-&-&-& \cr 
0&R^{21}_{12}&0&0&|&R^{21}_{21}&0&0&0&|&0&0&0&0&|&0&0&0&0& \cr 
0&0&0&0&|&0&R^{22}_{22}&0&0&|&0&0&0&0&|&0&0&0&0& \cr
0&0&0&R^{23}_{14}&|&0&0&R^{23}_{23}&0&|&0&R^{23}_{32}&0&0&|&R^{23}_{41}&0&0&0 
\cr
0&0&0&0&|&0&0&0&R^{24}_{24}&|&0&0&0&0&|&0&R^{24}_{42}&0&0 \cr
-&-&-&-&|&-&-&-&-&|&-&-&-&-&|&-&-&-&- \cr
0&0&R^{31}_{13}&0&|&0&0&0&0&|&R^{31}_{31}&0&0&0&|&0&0&0&0 \cr 
0&0&0&R^{32}_{14}&|&0&0&R^{32}_{23}&0&|&0&R^{32}_{32}&0&0&|&R^{32}_{41}&0&0&0 
\cr
0&0&0&0&|&0&0&0&0&|&0&0&R^{33}_{33}&0&|&0&0&0&0 \cr
0&0&0&0&|&0&0&0&0&|&0&0&0&R^{34}_{34}&|&0&0&R^{34}_{43}&0 \cr
-&-&-&-&|&-&-&-&-&|&-&-&-&-&|&-&-&-&- \cr
0&0&0&R^{41}_{14}&|&0&0&R^{41}_{23}&0&|&0&R^{41}_{32}&0&0&|&R^{41}_{41}&0&0&0 
\cr 
0&0&0&0&|&0&0&0&R^{42}_{24}&|&0&0&0&0&|&0&R^{42}_{42}&0&0 \cr
0&0&0&0&|&0&0&0&0&|&0&0&0&R^{43}_{34}&|&0&0&R^{43}_{43}&0 \cr
0&0&0&0&|&0&0&0&0&|&0&0&0&0&|&0&0&0&R^{44}_{44} \cr }   
\nn \eea      
\end{footnotesize}
where the non-zero entries are given by 
\bea 
R^{11}_{11}&=& \exp[(\m-\n)(\b-\d)]\frac{[u-\a][u-\a-1]}{[u+\a][u+\a+1]} \nn \\ 
R^{12}_{12}&=& \exp[(\m-\n)((1-k)\b-\d)]\frac{[u][u-\a-1]}{[u+\a][u+\a+1]}\nn\\
R^{13}_{13}&=& \exp[(\m-\n)(k\b-\d)]\frac{[u][u-\a-1]}{[u+\a][u+\a+1]} \nn \\
R^{14}_{14}&=& \exp[-(\m-\n)\d]\frac{[u][u-1]}{[u+\a][u+\a+1]} \nn \\
^*R^{12}_{21}&=& \exp[((1-k)\m-\n)(\b-\d)]q^u\frac{[\a][u-\a-1]}
                 {[u+\a][u+\a+1]}\nn \\
^*R^{13}_{31}&=& \exp[(k\m-\n)(\b-\d)]q^u\frac{[\a][u-\a-1]}{[u+\a][u+\a+1]}
                 \nn \\
R^{14}_{41}&=& \exp[-\n(\b-\d)]q^{2u}\frac{[\a][\a+1]}{[u+\a][u+\a+1]}\nn \\
^*R^{14}_{23}&=& \exp[(-k\n\b+((k-1)\m+\n)\d]
                 q^{u-1/2}\frac{[\a]^{1/2}[\a+1]^{1/2}[u]}{[u+\a][u+\a+1]}\nn\\
^*R^{14}_{32}&=& -\exp[(k-1)\n\b+(\n-k\m)\d]
                q^{u+1/2}\frac{[\a]^{1/2}[\a+1]^{1/2}[u]}{[u+\a][u+\a+1]}\nn\\
R^{21}_{21}&=& \exp[(\m-\n)(\b+(k-1)\d]\frac{[u][u-\a-1]}{[u+\a][u+\a+1]} \nn \\
^*R^{22}_{22}&=& \exp[(1-k)(\m-\n)(\b-\d)] \frac{[u-\a-1]}{[u+\a+1]}\nn \\
^*R^{23}_{23}&=& \exp[(\m-\n)(k\b+(k-1)\d)]\frac{[u]^2}{[u+\a][u+\a+1]}\nn \\
R^{24}_{24}&=& \exp[(k-1)(\m-\n)\d]\frac{[u]}{[u+\a+1]}\nn \\
R^{21}_{12}&=& -\exp[(\m+(k-1)\n)(\b-\d)]q^{-u}\frac{[\a][u-\a-1]}
                {[u+\a][u+\a+1]}\nn \\ 
^*R^{23}_{32}&=& \exp[(k\m+(k-1)\n)(\b-\d)]
                 \frac{2q-q^{2\a+1}-q^{-2\a-1}-q^{2u+1}+q^{2u-1}}
	       {(q^{u+\a}-q^{-u-\a})(q^{u+\a+1}-q^{-u-\a-1})}\nn \\
R^{24}_{42}&=& \exp[\n(k-1)(\b-\d)]q^u\frac{[\a+1]}{[u+\a+1]} \nn \\
^*R^{23}_{14}&=& -\exp[k\m\b+((1-k)\n-\m)\d]q^{-u+1/2}
                  \frac{[\a]^{1/2}[\a+1]^{1/2}[u]}{[u+\a][u+\a+1]}\nn\\
^*R^{23}_{41}&=& -\exp[(k\m-\n)\b+(1-k)\n\d]
                 q^{u+1/2}\frac{[\a]^{1/2}[\a+1]^{1/2}[u]}{[u+\a][u+\a+1]}\nn\\
R^{31}_{31}&=& \exp[(\m-\n)(\b-k\d)]\frac{[u][u-\a-1]}{[u+\a][u+\a+1]} \nn \\
^*R^{32}_{32}&=& \exp[(\m-\n)((1-k)\b-k\d)]\frac{[u]^2}{[u+\a][u+\a+1]}\nn \\ 
^*R^{33}_{33}&=& \exp[k(\m-\n)(\b-\d)] \frac{[u-\a-1]}{[u+\a+1]}\nn \\
R^{34}_{34}&=& \exp[-k(\m-\n)\d]\frac{[u]}{[u+\a+1]}\nn \\
R^{31}_{13}&=& -\exp[(\m-k\n)(\b-\d)]q^{-u}\frac{[\a][u-\a-1]}{[u+\a][u+\a+1]} 
                 \nn \\
^*R^{32}_{23}&=& \exp[(1-k)\m-k\n)(\b-\d)]
                 \frac{2q^{-1}-q^{2\a+1}-q^{-2\a-1}+q^{-2u+1}-q^{-2u-1}}
	       {(q^{u+\a}-q^{-u-\a})(q^{u+\a+1}-q^{-u-\a-1})} \nn \\
R^{34}_{43}&=& \exp[-k\n(\b-\d)]q^u\frac{[\a+1]}{[u+\a+1]}\nn \\
^*R^{32}_{14}&=& \exp[(1-k)\m\b-(\m-k\n)\d] 
                q^{-u-1/2}\frac{[\a]^{1/2}[\a+1]^{1/2}[u]}{[u+\a][u+\a+1]}\nn\\
^*R^{32}_{41}&=& \exp[((1-k)\m-\n)\b+k\n\d]
                q^{u-1/2}\frac{[\a]^{1/2}[\a+1]^{1/2}[u]}{[u+\a][u+\a+1]}\nn\\
R^{41}_{41}&=& \exp[(\m-\n)\b]\frac{[u][u-1]}{[u+\a][u+\a+1]}\nn \\
R^{42}_{42}&=& \exp[(1-k)(\m-\n)\b]\frac{[u]}{[u+\a+1]}\nn \\
R^{43}_{43}&=& \exp[k(\m-\n)\b]\frac{[u]}{[u+\a+1]}\nn \\
R^{44}_{44}&=& 1\nn \\
R^{41}_{14}&=& \exp[\m(\b-\d)]q^{-2u}\frac{[\a][\a+1]}{[u+\a][u+\a+1]}\nn \\
^*R^{42}_{24}&=& -\exp[\m(1-k)(\b-\d)]q^{-u}\frac{[\a+1]}{[u+\a+1]}\nn \\
^*R^{43}_{34}&=& -\exp[k\m(\b-\d)]q^{-u}\frac{[\a+1]}{[u+\a+1]}\nn \\
^*R^{41}_{23}&=& \exp[(\m-k\n)\b+(k-1)\m\d]
               q^{-u-1/2}\frac{[\a]^{1/2}[\a+1]^{1/2}[u]}{[u+\a][u+\a+1]}\nn\\  
^*R^{41}_{32}&=& -\exp[(\m+(k-1)\n)\b-k\m\d]
                 q^{-u+1/2}\frac{[\a]^{1/2}[\a+1]^{1/2}[u]}
	       {[u+\a][u+\a+1]}.\nn      \eea  

We remark again that the above $R$-matrix solves the MYB
equation subject to the rule (\ref{rule}) which is a natural consequence 
of the superalgebra structure underlying the $R$-matrix. However, 
a change of signs in those matrix elements with a $*$ gives a  solution
which satisfies the Yang-Baxter equation in the usual matrix form. 

The four dimensional space of states associated with the $U_q(sl(2|1))$
representation may be identified with the electronic states
$$\left|0\right>,~~\left|\uparrow\right>,~~\left|\downarrow\right>,~~
\left|\uparrow\downarrow\right>. $$ 
Taking the logarithmic derivative of the transfer matrix yields the following
global Hamiltonian on a periodic lattice (with convenient normalization)
\bea H&=&-\sum_{i,\sigma}(c_{i\sigma}^{\dagger}c_{(i+1)\sigma} +h.c.)
\exp[-1/2(\eta-\sigma\epsilon)n_{i(-\sigma)}-1/2(\eta+\sigma\epsilon)
n_{(i+1)(-\sigma)}] \nn \\
&&+\frac{U}{2}\sum_i(c^{\dagger}_{i\uparrow}c^{\dagger}_{i\downarrow}
c_{(i+1)\downarrow}c_{(i+1)\uparrow}+h.c.) +U\sum_i n_{i\uparrow}n_{i 
\downarrow}  \nn \\
&&+ (q^{\a+1}+q^{-\a-1}-\frac{(\m-\n)(q^{\a+1}-q^{-\a-1})}{4\ln q})\sum_i 
n_i \nn \\
&&-\frac{(\m-\n)(1-2k)(q^{\a+1}-q^{-\a-1})}{2\ln q}\sum_i
S^z_i   \nn \eea 
where 
$$\exp\epsilon=q^{-1},~~~~\exp(-\eta)=\frac{[\a+1]}{[\a]},~~~~U
=2[\a]^{-1}$$
and the standard notation has been used for the Fermi and spin operators. 
The above model is the supersymmetric $U$ model with  
chemical potetntial and applied magnetic field whose couplings may 
be chosen arbitrarily through the parameters $\m-\n$ and $k$.

\vfil\eject

\centerline{\bf 6. The Bariev model}
~~\\
The Bariev model on a one-dimensional periodic lattice has the Hamiltonian
$$H=-\sum_{i,\sigma}(c_{i\sigma}^{\dagger}c_{(i+1)\sigma} +h.c.)
\exp[-1/2(\eta-\sigma\eta)n_{i(-\sigma)}-1/2(\eta+\sigma\eta)
n_{(i+1)(-\sigma)}]. $$
It is apparent that the Bariev model is obtainable from the 
supersymmetric $U$ model above in the limit $\a\rightarrow\infty$   
by choosing $\m-\n=4\ln q$ and $k=1/2$. 
It would thus appear 
reasonable to suspect that an $R$-matrix solution for the Bariev model is 
obtainable from the solution for the supersymmetric $U$ model. However, if one
is to naively take the limit $\a\rightarrow\infty$ one finds
$$\lim_{\a\rightarrow\infty} R(u,\b,\d)=M^{\n(\d-\b)}_1M_2^{\m(\b-\d)}P$$ 
and as such the Bariev model is not obtainable in this fashion.  

Alternatively, we may make the following rescaling of the parameters
\bea u\rightarrow\a u, &&~~~~v\rightarrow\a v, \nn \\
\m\rightarrow 2\m\ln q, &&~~~~\n\rightarrow 2\n \ln q, \nn \\
q\rightarrow q^{1/\a}   &&  \nn \eea 
and set 
$$\b=u,~~~~\d=v,~~~~k=1/2.$$ 
In this manner we obtain a solution of the MYB equation 
$$R_{12}(u,v)R_{13}(u,w)R_{23}(v,w)=R_{23}(v,w)R_{13}(u,w)R_{12}(u,v)$$
in the limit $\a\rightarrow\infty $ where the matrix elements of $R(u,v)$ 
now read
\bea 
R^{11}_{11}&=& q^{2(\m-\n)(u-v)}\frac{[u-v-1]^2}{[u-v+1]^2} \nn \\
R^{12}_{12}&=& q^{(\m-\n)(u-2v)}\frac{[u-v][u-v-1]}{[u-v+1]^2}\nn\\
R^{13}_{13}&=& q^{(\m-\n)(u-2v)}\frac{[u-v][u-v-1]}{[u-v+1]^2} \nn \\
R^{14}_{14}&=& q^{-2(\m-\n)v}\frac{[u-v]^2}{[u-v+1]^2} \nn \\
^*R^{12}_{21}&=& q^{(\m-2\n+1)(u-v)}\frac{[u-v-1]}
                 {[u-v+1]^2}\nn \\
^*R^{13}_{31}&=& q^{(\m-2\n+1)(u-v)}\frac{[u-v-1]}{[u-v+1]^2}
                 \nn \\
R^{14}_{41}&=& q^{(2-2\n)(u-v)}\frac{1}{[u-v+1]^2}\nn \\
^*R^{14}_{23}&=& q^{(1-\n) u+(-\m+2\n-1)v}
                 \frac{[u-v]}{[u-v+1]^2}\nn\\
^*R^{14}_{32}&=& -q^{(1-\n) u+(2\n-\m-1)v}
                \frac{[u-v]}{[u-v+1]^2}\nn\\
R^{21}_{21}&=& q^{(\m-\n)(2u-v)}\frac{[u-v][u-v-1]}{[u-v+1]^2} \nn \\
^*R^{22}_{22}&=& q^{(\m-\n)(u-v)} \frac{[u-v-1]}{[u-v+1]}\nn \\
^*R^{23}_{23}&=& q^{(\m-\n)(u-v)}\frac{[u-v]^2}{[u-v+1]^2}\nn \\
R^{24}_{24}&=& q^{-(\m-\n)v}\frac{[u-v]}{[u-v+1]}\nn \\
R^{21}_{12}&=& -q^{(2\m-\n-1)(u-v)}\frac{[u-v-1]}
                {[u-v+1]^2}\nn \\
^*R^{23}_{32}&=& -q^{(\m-\n)(u-v)} \frac{1}{[u-v+1]^2}\nn \\
R^{24}_{42}&=& q^{(1-\n)(u-v)}\frac{1}{[u-v+1]} \nn \\
^*R^{23}_{14}&=& -q^{(\m-1) u+(\n-2\m+1)v}
                  \frac{[u-v]}{[u-v+1]^2}\nn\\
^*R^{23}_{41}&=& -q^{(\m-2\n+1)u+(\n-1) v}
                 \frac{[u-v]}{[u-v+1]^2}\nn\\
R^{31}_{31}&=& q^{(\m-\n)(2u-v)}\frac{[u-v][u-v-1]}{[u-v+1]^2} \nn \\
^*R^{32}_{32}&=& q^{(\m-\n)(u-v)}\frac{[u-v]^2}{[u-v+1]^2}\nn \\
^*R^{33}_{33}&=& q^{(\m-\n)(u-v)} \frac{[u-v-1]}{[u-v+1]}\nn \\
R^{34}_{34}&=& q^{-(\m-\n)v}\frac{[u-v]}{[u-v+1]}\nn \\
R^{31}_{13}&=& -q^{(2\m-\n-1)(u-v)}\frac{[u-v-1]}{[u-v+1]^2}
                 \nn \\
^*R^{32}_{23}&=& -q^{(\m-\n)(u-v)} \frac{1}{[u-v+1]^2} \nn \\
R^{34}_{43}&=& q^{(1-\n)(u-v)}\frac{1}{[u-v+1]}\nn \\
^*R^{32}_{14}&=& q^{(\m-1) u+(\n-2\m+1)v}
                \frac{[u-v]}{[u-v+1]^2}\nn\\
^*R^{32}_{41}&=& q^{(1+\m-2\n)u+(\n-1) v}
                \frac{[u-v]}{[u-v+1]^2}\nn\\
R^{41}_{41}&=& q^{2(\m-\n)u}\frac{[u-v]^2}{[u-v+1]^2}\nn \\
R^{42}_{42}&=& q^{(\m-\n)u}\frac{[u-v]}{[u-v+1]}\nn \\
R^{43}_{43}&=& q^{(\m-\n)u}\frac{[u-v]}{[u-v+1]}\nn \\
R^{44}_{44}&=& 1\nn \\
R^{41}_{14}&=& q^{(2\m-2)(u-v)}\frac{1}{[u-v+1]^2}\nn \\
^*R^{42}_{24}&=& -q^{(\m-1)(u-v)}\frac{1}{[u-v+1]}\nn \\
^*R^{43}_{34}&=& -q^{(\m-1)(u-v)}\frac{1}{[u-v+1]}\nn \\
^*R^{41}_{23}&=& q^{(2\m-\n-1) u+(1-\m) v}
               \frac{[u-v]}{[u-v+1]^2}\nn\\
^*R^{41}_{32}&=& -q^{(2\m-\n-1)u+(1-\m) v} 
                  \frac{[u-v]}{[u-v+1]^2}.\nn      \eea
However, calculating the Hamiltonian associated with this solution one finds
$$H=-\sum_{i,\sigma}(c_{i\sigma}^{\dagger}c_{(i+1)\sigma} +h.c.)
+\left[1/2(q-q^{-1})(\n-\m)+(q+q^{-1})\right]\sum_in_i$$ 
which of course is simply a free fermion model with chemical potential
and corresponds to a particular case of the Bariev model.

\centerline{{\bf 7. Conclusions}}
~~\\
Here it has been demonstrated how $R$-matrices which satisfy the MYB equation
such that they do not have the difference property may be obtained. As an 
example, the anisotropic supersymmetric $U$ model with arbitrary chemical 
potential and magnetic field terms was derived, which, in a particular limit,
reproduces the Bariev model. 
Regrettably, this approach does not shine a light on the connection 
with the $R$-matrix 
solutions  
for the Bariev model computed in \cite{z,sw} without the difference property.
It is of course possible that a different limiting procedure may yield an
$R$-matrix solution for the Bariev model but its form is not apparent at 
present.

An open problem for future work is to investigate whether Shastry's 
solution \cite{s} and the recently introduced $SU(N)$ Hubbard models
\cite{mas,mjm} may be related to an $R$-matrix with the difference 
property via a twisting procedure.

~~\\
~~\\
\centerline{{\bf Acknowledgements}}
~~\\

This work is supported by the Australian Research Council. Many thanks to A. 
Foerster for assistance, M.J. Martins for discussions which motivated me
to look at this problem and Y.-Z. Zhang for proofreading the manuscript.


\end{document}